\documentclass[%
aps,superscriptaddress,  
nofootinbib, %
reprint,
a4paper,
longbibliography
]{revtex4-2} %

\newif\ifarxiv 

\usepackage{graphicx}
\usepackage{float}
\usepackage{amsfonts,amstext}
\usepackage{url}
\usepackage{xcolor}
\usepackage{braket}
\usepackage{soul}
\usepackage{amsthm}
\usepackage{physics}
\usepackage{hhline}
\usepackage{bbm}
\usepackage{enumerate}
\usepackage{comment}
\usepackage{colortbl}
\usepackage{amssymb}
\usepackage{comment}

\AtBeginDocument{%
    \newwrite\bibnotes
    \def\bibnotesext{Notes.bib}
    \immediate\openout\bibnotes=\jobname\bibnotesext
    \immediate\write\bibnotes{@CONTROL{REVTEX41Control}}
    \immediate\write\bibnotes{@CONTROL{%
    apsrev41Control,author="08",editor="1",pages="1",title="0",year="0"}}
     \if@filesw
     \immediate\write\@auxout{\string\citation{apsrev41Control}}%
    \fi
  }%

\usepackage[nodayofweek]{datetime}  
\usepackage{hyperref}
\definecolor{mylinkcolor}{rgb}{0,0,0.8} 
\hypersetup{unicode=true, %
  bookmarksnumbered=false,bookmarksopen=false,bookmarksopenlevel=1, %
  breaklinks=true,pdfborder={0 0 0},colorlinks=true}%
\hypersetup{%
  anchorcolor=mylinkcolor,citecolor=mylinkcolor, %
  filecolor=mylinkcolor,linkcolor=mylinkcolor, %
  menucolor=mylinkcolor,runcolor=mylinkcolor, %
  urlcolor=mylinkcolor}%

\makeatletter 
    
\renewcommand\onecolumngrid{
\do@columngrid{one}{\@ne}%
\def\set@footnotewidth{\onecolumngrid}
\def\footnoterule{\kern-6pt\hrule width 1.5in\kern6pt}%
}

\renewcommand\twocolumngrid{
        \def\footnoterule{
        \dimen@\skip\footins\divide\dimen@\thr@@
        \kern-\dimen@\hrule width.5in\kern\dimen@}
        \do@columngrid{mlt}{\tw@}
}%

\makeatother  

\renewcommand{\emph}[1]{\textit{#1}}

\newcommand{\chanfin}[1]{{#1}} 

\newtheoremstyle{theorem}
	{6pt}
	{}
	{\itshape}
	{}
	{\bfseries}
	{:}
	{.5em}
	{}
\theoremstyle{theorem}

\newtheorem{example}{Example}[section]

\newtheoremstyle{defn}
	{6pt}
	{}
	{}
	{}
	{\bfseries}
	{:}
	{.5em}
	{}
\theoremstyle{defn}
\newtheorem{defn}{Definition}[section]

\newtheoremstyle{red}
{6pt}
{}
{\color{red}\itshape}
{}
{\color{red}\bfseries}
{:}
{.5em}
{}
\theoremstyle{red}

\begin{document}

\title{Monogamy relations for relativistically causal correlations}

\author{Mirjam Weilenmann}
\affiliation{Département de Physique Appliquée, Université de Genève, Genève, Switzerland}
\email{mirjam.weilenmann@unige.ch}

\date{\today}

\begin{abstract}
Non-signalling conditions encode minimal requirements that any (quantum) systems must satisfy in order to be consistent with special relativity. Recent works have argued that in scenarios involving more that two parties, { correlations} compatible with relativistic causality do not have to satisfy all possible non-signalling conditions but only a subset of them. Here we show that correlations satisfying  only this subset of constraints have to satisfy highly non-local monogamy relations between the effects of space-like separated random variables. These monogamy relations take the form of entropic inequalities between the various systems and we give a general method to derive them. Using these monogamy relations we refute previous suggestions for physical mechanisms that could lead to relativistically causal correlations, demonstrating that such mechanisms would lead to superluminal signalling.
\end{abstract}

\maketitle

\section{Introduction}
The relation between quantum and relativity theory has been a topic of interest in the realm of quantum information theory in the last few decades. Investigations into whether consistency with special relativity is sufficient to restrict correlations between two parties to be quantum~\cite{PR}, has prompted extensive research in quantum foundations into the physical principles underlying it. At the same time, cryptographic tasks have been analysed against adversaries that are not necessarily restricted to act according to  quantum theory but instead are restricted by relativistic principles { captured by \emph{non-signalling constraints}; 
 examples of such tasks include} device-independent key distribution~\cite{qkd_nonsignalling} and randomness extraction~\cite{hanggi, randomness_recent}.

As quantum technologies progress to involve larger systems { and} more parties, e.g. in quantum networks, the question of 
{ generalising these concepts} becomes increasingly important. 
It was argued early on that the straightforward generalisation of non-signalling { to more parties}~\cite{Masanes}
can be relaxed in specific setups. That is, if such a relaxation does not introduce causal inconsistencies such as causal loops~\cite{shimony, GPR_jam}. 
This led to a general notion of \emph{relativistically causal correlations}~\cite{HR_jam}.

Subsequent work showed that information-theoretic and statistical challenges may arise when restricting correlations in the multi-party setting only by this notion of relativistic causality. Specifically, it leads to no-go theorems for cryptographic protocols~\cite{HR_jam2} as well as to challenges for causal modelling, which in this case relies on fine-tuning causal dependencies~\cite{VilasiniRogerPRL, VilasiniRogerPRA}.

{ Here}, we demonstrate conceptual implications of allowing the most general relativistically causal correlations. While the examples considered in~\cite{shimony,GPR_jam,HR_jam,HR_jam2} seem to suggest that causality is relaxed in a somewhat innocent way, we show that in more complicated multi-party setups, these correlations have to satisfy a new type of monogamy relation (Section~\ref{sec:monogamy}). The special feature of these relations is that they relate the influences { that} independent, space-like separated variables { can have}. We argue that this type of correlation is highly non-local {in a new way}.

Furthermore, {correlations alone do
not explain, what mechanisms they might arise from.} 
For understanding in what sense correlations can be generated in an information theoretic task or used by an adversary to tamper with a cryptographic protocol, { these mechanisms} and the control different parties may have over them are, however, important. Despite the lack of a full explanation, a partial intuition of such a mechanism was given in~\cite{GPR_jam}, where there is a ``jammer'', \emph{who has access to a
jamming device which he can activate, at will}~\cite{GPR_jam}. 

We show here, that our monogamy relations can be used to rule out the explanation of correlations in terms of a mechanism that an agent can {(partially)} control or activate, as such { control} would allow for a superluminal transfer of information. Thus the above intuition is inadequate in \chanfin{the more} general \chanfin{scenarios we consider}. We go on to discuss further problems the monogamy relations create for other potential explanations of such a mechanism as well as for causal modelling. This work thus restricts the type of possible explanations of relativistically causal correlations, \chanfin{calling into question whether} these correlations will be realised by a physical theory (Section~\ref{sec:signalling}).

Our monogamy relations can be phrased in terms of entropic inequalities for which we propose a technique to derive them. We further show how this technique can be adapted to relax entropic constraints on non-signalling correlations to their relativistically causal { analogues}. We illustrate this with an example showing how the tradeoff between the bipartite non-locality that can be shared in a three-party setting between pairs of parties is altered in this paradigm as compared to imposing all non-signalling constraints~\cite{BudroniChaves} (Section~\ref{sec:crypto}).

\section{Results}

\begin{figure}
    \centering
    \includegraphics[trim={3.7cm 5.6cm 14cm 4.2cm},clip,width=\columnwidth]{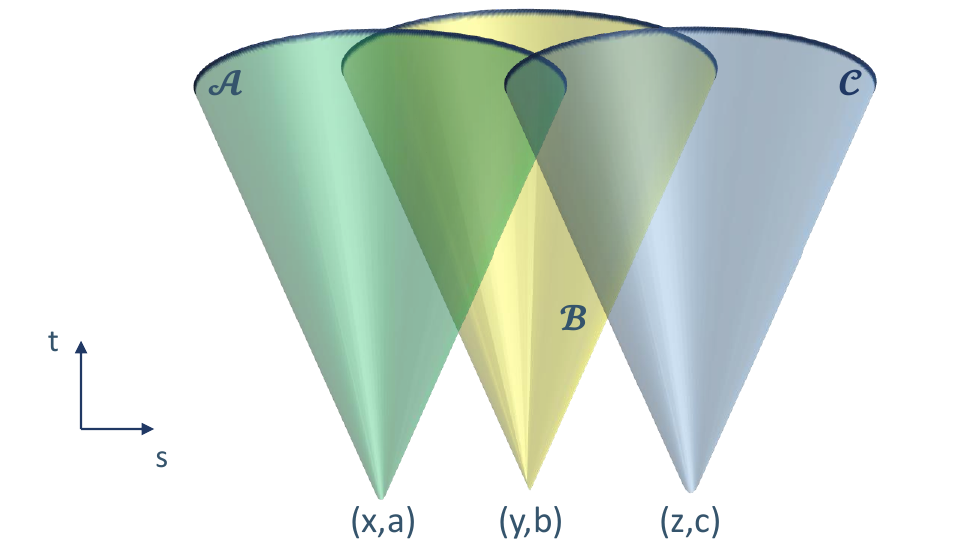}
    \caption{{\bf Contrasting multi-partite generalisations of non-signalling conditions.} Three-party scenario involving Alice, Bob and Charlie with inputs (outputs) $x, (a)$, $y, (b)$ and $z, (c)$ respectively. The parties are arranged on a line and such that their choice of input and generation of output happens space-like separated, as illustrated with the three future light cones. As the intersection of the light cones $\mathcal{K}_{(x,a)}$ and $\mathcal{K}_{(z,c)}$ lies fully within $\mathcal{K}_{(y,b)}$, { the relativistic causality constraints do not include} $P(ac|xyz)=P(ac|xz)$ \chanfin{ $\forall \ a,c,x,y,z$}, which is among the { usual non-signalling constraints~\cite{Masanes}}. Notice that this means that Bob's input can only affect the correlations of Alice and Charlie not their marginals, { i.e., $P(a|xyz)=P(a|x)$, $P(c|xyz)=P(c|z)$\chanfin{ $ \ \forall \ a,c,x,y,z$}.} { One way to achieve such an influence is by means of} the  \chanfin{functional dependency} $A\oplus C =Y$,  { which will be used throughout this work.} 
    }
    \label{fig:bob_jams}
\end{figure}

{ The present work is concerned with the constraints imposed by relativistic causality on the correlations observed by multiple parties~\cite{shimony, GPR_jam, HR_jam}.} 
{ We mean by \emph{relativistic causality} the requirement} that no observer can send superluminal signals, in the sense that an input $X$ can influence correlations of a set of variables $\{A_i\}_i$, if and only if the intersection \chanfin{$\cap_i  \mathcal{K}_{a_i}$} of the future light cones \chanfin{$\mathcal{K}_{a_i}$} of these variables lies fully in the future light cone of $X$, otherwise it cannot, i.e., otherwise $P(\{a_i\}_i|x)=P(\{a_i\}_i)$ \chanfin{ $\forall \{a_i\}_i,x$.} This arguably corresponds to the notion captured by the mathematical formalisation from~\cite{GPR_jam, HR_jam, HR_jam2}.\footnote{{In this work we choose to formulate the framework of relativistic causality directly via the mathematical constraints. This is because while} relativistic causality is { sometimes} introduced  as a setup where no causal loops occur~\cite{HR_jam}, it was { previously} pointed out~\cite{VilasiniRogerPRL} that this is not -- as claimed -- in 1:1 correspondence with the mathematical constraints. { The formulation of the mathematical constraints, on the other hand, \chanfin{is so far} uncontroversial.} 
}
The standard tripartite example comparing these constraints to the full non-signalling constraints was previously considered in~\cite{GPR_jam, HR_jam}, and is presented in Figure~\ref{fig:bob_jams}.
The notion of relativistic causality captures the intuition that having variables affect correlations of other variables does not lead to logical contradictions { whenever these correlations can only be evaluated in the \chanfin{future of the former variables}}~\cite{GPR_jam, HR_jam}. 
In the example of Figure~\ref{fig:bob_jams} this is the case for the variables $A$ and $C$, the correlations of which can only be seen in the future of $Y$.

{ Formally,} we define the constraints imposed by relativistic causality as follows. 
\begin{defn}[Relativistic causality constraints]~\label{def:rel_caus} 
{
Let there be $n$-parties labelled $1, \ldots, n$ and let there be a strict subset of the parties $ \mathcal{J} \subset \{1, \ldots, n\}$ and let $\mathcal{J}^c = \{1, \ldots, n\} \setminus \mathcal{J}$. Let  $\mathcal{A}_\mathcal{J}=\{a_i \mid i \in \mathcal{J} \}$ and $\mathcal{X}_\mathcal{J}=\{x_i \mid i \in \mathcal{J} \}$ and define $\mathcal{X}_\mathcal{K}$ to be the set of $x_k \in \mathcal{X}_{\mathcal{J}^c}$ such that the intersection of the future of all the variables in  $A_\mathcal{J}$,
$\cap_{j \in \mathcal{J}}  \mathcal{K}_{a_j}$, lies within the future of $X_k$. Then, $\forall \ \mathcal{J}$, 
$$\sum_{a_i \, :  \, i \in \mathcal{J}^c} P(a_1, \ldots, a_n|x_1, \ldots, x_n) = P(\mathcal{A}_\mathcal{J}| \mathcal{X}_\mathcal{J}\mathcal{X}_\mathcal{K}),$$}
\chanfin{$\forall \ a_1, \ldots, a_n, x_1, \ldots, x_n$}
\end{defn} 
Notice that scenarios where only a subset of the involved parties are considered are obtained by marginalising over the other parties (outcomes). In the following we call correlations that satisfy all relativistic causality constraints in a light-cone arrangement the \emph{relativistically causal correlations} of that scenario.

While this formal definition is stated straightforwardly, the physical notion behind it { deserves a few additional remarks.} 
{ Firstly, relativistically causal correlations differ from general multipartite non-signalling correlations~\cite{Masanes} (see Figure~\ref{fig:bob_jams} for an example).}

{ Secondly,} this notion is incompatible with the operational reasoning that is { usually} followed in quantum information theory (or in treatments in quantum foundations, like generalised probabilistic theories~\cite{Hardy2001, Barrett2005}), where the non-signalling conditions are naturally built into the { operational} framework.  In the quantum case this { is done} by taking space-like separated measurements to be made up from commuting operators or to act on different Hilbert spaces in a tensor product. However, { it was suggested that relativistically causal correlations} might appear in future physical theories~\cite{HR_jam}. 

{ Thirdly, the relativistic causality constraints are  frame-independent, because} any event in the future light cone of some event is in this future light cone in any reference frame. The analogous statement holds for events in the intersection of several light cones.  
We discuss this { as well as the dependency of the relativistic causality constraints on the number of spatial dimensions of the setup in { Supplementary Note 1.}}

{ While the set of relativistically causal correlations is thus straightforwardly defined,}
an underlying explanation, i.e., a mechanism leading to such correlations is not known and left open { in previous work}~\cite{GPR_jam, HR_jam}. 
The intuition that is given is that an agent can activate a device that ``jams'' correlations of other parties~\cite{GPR_jam}. While in~\cite{GPR_jam} this mechanism is instantaneous, it is considered in terms of superluminal influences that travel at finite speeds and considered among a larger number of parties in~\cite{HR_jam}.

Note that while the { relativistic causality constraints} remain valid  independently of the frame, these underlying explanations \emph{depend on the reference frame}, as the condition that a signal from a space-time point $Z$ travelling at a speed $u>c$ can reach a point $A$ is not Lorentz invariant. Indeed, if a signal can travel this way in one reference frame, there may be another reference frame where the order of events at $Z$ and $A$ is reversed. Thus, such an explanation can only be retained in certain reference frames~\cite{GPR_jam} and may be different in others. 

{
Notice that these explanations specifically aim to explain the mechanism that leads to superluminal influences beyond standard non-signalling. For this, an explanation in the sense of the quantum formalism for quantum correlations (or a generalised probabilistic theory for non-signalling correlations) is not strong enough (as both abide by non-signalling). Thus an explanation or mechanism for these correlations has to be of a new, more general nature; this might involve some effect that propagates this influence  (in the spirit of the "jamming" intuition above) or might have to be of a new, as yet inconceivable nature. 

Finding such a mechanism is thus a general problem, that differs from attempting to replace the usual quantum  (or generalised probabilistic) explanation of quantum (or non-signalling) correlations by means of hidden communication~\cite{HC1, HC2, HC3, HC4, HC5}. One could, however, attempt to use features in the spirit of a hidden communication mechanism in combination with existing theories to explain relativistically causal correlations. 
\chanfin{However,} the set of relativistically causal correlations \chanfin{cannot} arise solely from hidden communication models where influences propagate at finite speeds~\cite{HC1,HC2,HC3,HC4,HC5,HR_jam}.
}

{ In the following subsections we present our results for relativistically causal correlations.}

\subsection{Monogamy of relativistically causal correlations} \label{sec:monogamy}
\begin{figure}
    \centering
    \includegraphics[trim={2.2cm 1.2cm 1.4cm 1.0cm},clip,width=\columnwidth]{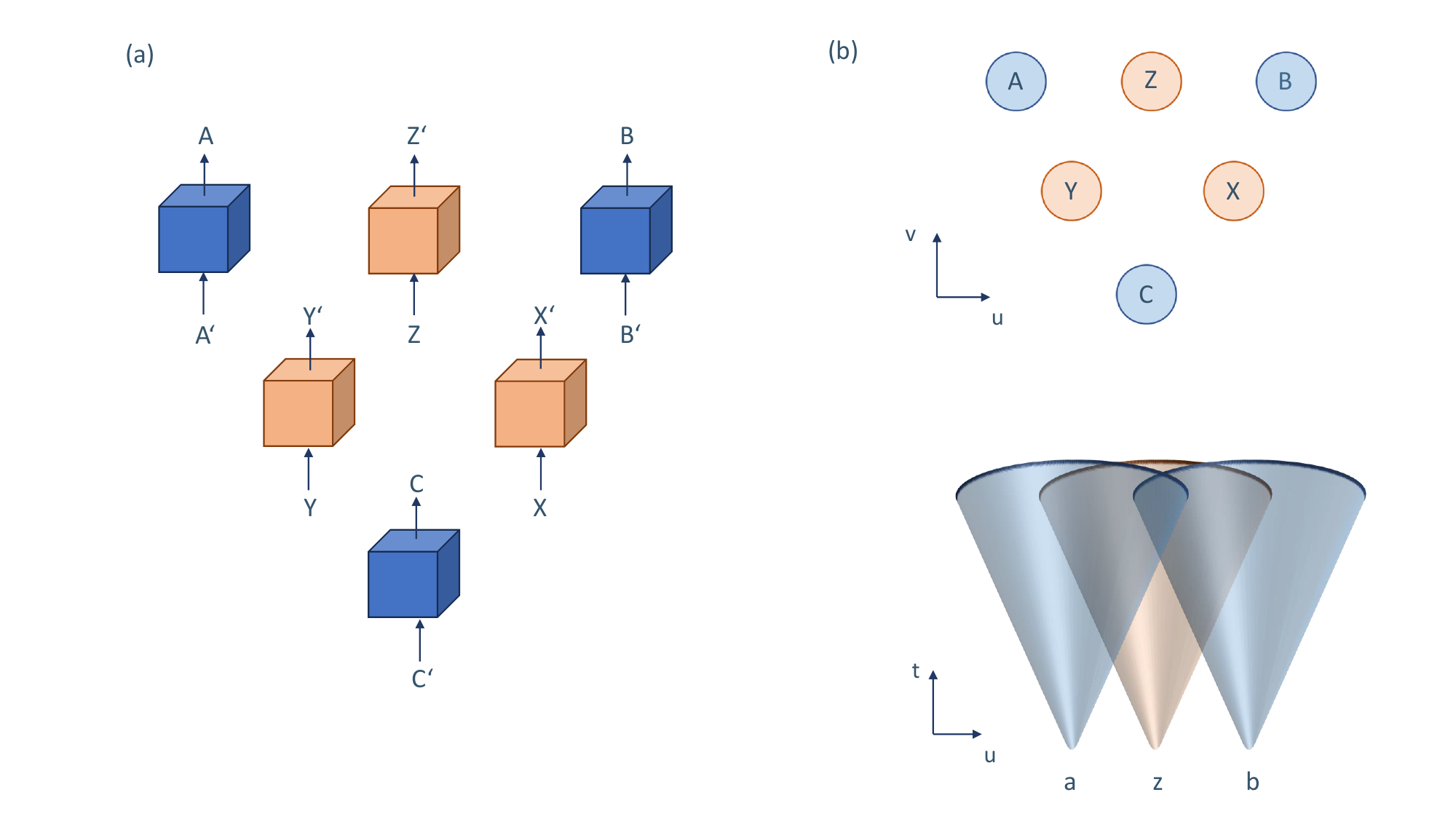}
    \caption{{\bf Scenario in multiple spatial dimensions.} { The three parties in orange can choose inputs $X$, $Y$, $Z$ and obtain outcomes $X'$, $Y'$, $Z'$ while Alice, Bob and Charlie choose inputs $A'$, $B'$, $C'$ and obtain outcomes $A$, $B$ and $C$ respectively. For the correlations considered in the main text, only $A,B,C,X,Y,Z$ play a role.} The six events { where these variables are generated} are space-like separated and are displayed here in a frame where they are simultaneous { (see (b)).} 
    }
    \label{fig:structure}
\end{figure}

In this section, we show that for scenarios with more than three parties, the relativistically causal correlations have to satisfy strong monogamy constraints. { We show this  by means of the setup in Figure~\ref{fig:structure} and refer to the {Supplementary Information} for other examples and  \chanfin{to the Methods} for a general method to formulate entropic monogamy inequalities.}

{ Thus, let us consider the setting illustrated} in Figure~\ref{fig:structure}. { This scenario} allows us to isolate the effect of non-local relativistically causal influences from local ones by denoting only either input or outcome at each space-time point and ignoring the other. This means that correlations are solely established via non-local effects.
In this scenario, the light-cone arrangement imposes, according to Definition~\ref{def:rel_caus}, the relations
\begin{align}
	P(ab|xyz) &= P(ab|z) \nonumber\\
	P(ac|xyz) &= P(ac|y) \nonumber\\
	P(bc|xyz) &= P(bc|x) \nonumber\\
	P(a|xyz) &= 	P(a) \label{eq:jam}\\
	P(b|xyz) &= 	P(b) \nonumber\\
	P(c|xyz) &= 	P(c) , \nonumber
\end{align}
\chanfin{$\forall \ a,b,c,x,y,z$}, which give rise to the set of relativistically causal correlations
\begin{align*} 
\mathcal{K}= \left\{ \chanfin{{P}} \mid \right. &\left.P(abc|xyz) \geq 0 \ \forall a,b,c,x,y,z, \right. \\ 
&\left.\sum_{a, b, c} P(abc|xyz) = 1  \ \forall x,y,z \right. \\
&\left.\text{and the equalities } \eqref{eq:jam} \text{ hold }\right\}.
\end{align*}
Performing a vertex enumeration, we can express any correlations  $\chanfin{{P}} \in \mathcal{K}$ as a convex combination of the finite number of extremal vertices of the polytope $\mathcal{K}$. For the case of binary $a,b,c,x,y,z$, we have performed this computation with the software PORTA~\cite{PORTA}. This leads to a set of correlations that is spanned by 11964 extremal vertices. Out of these we find that of the three constraints
\begin{align}
P(ab|xyz) =P(ab), \label{eq:j1} \\
P(ac|xyz) =P(ac), \label{eq:j2}\\
P(bc|xyz) =P(bc), \label{eq:j3}
\end{align}
\chanfin{ $\forall \ a,b,c,x,y,z$} which have been dropped compared to { the standard non-signalling constraints~\cite{Masanes} (see also Supplementary Equation~(S1))}, $8624$ extremal vertices violate all three, $3 \cdot 968$ violate two of them, $3 \cdot 60$ violate one and $256$ violate none. Due to the large number of extremal vertices we do not list them here.

An example of a compatible distribution is constructed using the following functional dependency of variables: In this setup, it is possible that $Z$ determines whether $A$ and $B$ are correlated or anti-correlated, i.e.\, $Z=A \oplus B$. This in turn implies that $X$ and $Y$ cannot be causally connected to the correlations of $B, C$ or $A, C$ respectively. Indeed, combining the constraints specifying $\mathcal{K}$ and $Z=A \oplus B$ for the marginal $P(ab|z)$, we find that all compatible distributions satisfy  $P(abc|xyz)=P(abc|z)$ \chanfin{ $\forall \ a,b,c,x,y,z$}. We derive this by running  again a vertex enumeration for this scenario and checking that this holds for all extremal vertices. Indeed, only few extremal correlations are left in this case, which we list in { Supplementary Note 2},  
all of which satisfy \eqref{eq:j2} and \eqref{eq:j3}. This implies that there is a \emph{monogamy relation} between these relativistically causal influences, where the effect of $Z$ on $A,B$ restricts possible influences of the space-like separated $X$ and $Y$. 

We remark here that monogamy relations between the influences of $X,Y,Z$ are not restricted to this particular example but hold more generally; { the} example is chosen for ease of presentation. In { Supplementary Note 2}, we also show monogamy for other functional dependencies of the variables in this setup. Specifically, we consider the setup of Figure~\ref{fig:structure} where each of the six parties chooses inputs and obtains outcomes.
\chanfin{We then show that if 
correlations}  previously considered in~\cite{HR_jam} are established in a sub-scenario of the setup, they imply monogamy relations \chanfin{in the sense that they prevent other parties} from achieving the same \chanfin{dependencies}. 

We further derive  general constraints on the information the  variables $X,Y,Z$ can hold about the pairs of variables $(B,C), (A,C), (A,B)$ respectively in the setup of Figure~\ref{fig:structure}.  A natural measure for capturing this is the \emph{mutual information} of two random variables $S,T$, $I(S:T)=H(S)+H(T)-H(ST)$, where $H(S)= -\sum_{s} P(s) \log P(s)$ is the Shannon entropy of the random variable $S$. The mutual information is { non-negative} {and} it is zero iff $P(st)=P(s)P(t)\forall s,t$, thus indeed capturing the information $T$ holds about $S$ and vice versa. We will further use the \emph{conditional mutual information} of two random variables $S,T$ conditioned on a third, $V$, defined as 
$I(S:T|V)=H(SV)+H(TV)-H(V)-H(STV)$, which is well-known to be { non-negative} for any sets of random variables $S,T,V$ and zero iff $P(st|v)=P(s|v)P(t|v) \ \forall s,t,v$.

Let us first consider random variables $A,B,C,X,Y,Z$ with restricted cardinality, again in the setup of Figure~\ref{fig:structure}. 
For $X$, $Y$, $Z$ uniformly random bits, and binary $A$, $B$, $C$ we find that 
\begin{equation}
I(AB:Z)+I(AC:Y)+I(BC:X) \leq 1. \label{eq:bit_monogamy}
\end{equation}
This is a non-trivial restriction on the $P(abcxyz)$, as each of the terms $I(AB:Z),I(AC:Y),I(BC:X)$ can separately take the value $1$. To prove the inequality we {can} check that it holds for all of the 11964 extremal vertices of $\mathcal{K}$ (see above).   Furthermore, $P(x), P(y), P(z)$ respectively are fixed in this case and we know that the mutual information is a convex function in $P(bc|x), P(ac|y), P(ab|z)$ respectively, which concludes the proof, since  the convex combination of two distributions $P_1$, $P_2$, $\lambda P_1(abcxyz) +(1-\lambda) P_2(abcxyz)$ in this case corresponds to a convex combination $\lambda P_1(ab|z) +(1-\lambda) P_2(ab|z)$, $\lambda P_1(ac|y) +(1-\lambda) P_2(ac|y)$, $\lambda P_1(bc|x) +(1-\lambda) P_2(bc|x)$ and a convex function on a convex set takes its maximal value at an extreme point, which in a polytope is at a vertex.

This entropy inequality captures the monogamy between the influences of $X,Y,Z$: it ensures that if $Z$ has maximal information about $A,B$, which in this case means $I(Z:AB)=1$, then $I(Y:AC)=I(X:BC)=0$, which is the case if and only if $Y$ is independent of $AC$ and $X$ is independent of $AB$. The strategy $Z=A\oplus B$ above is an example that saturates this inequality.

We can further derive constraints that hold independently of the cardinality of the involved variables (and without assuming uniform $X,Y,Z$): 
\begin{equation}
I(X:BC)+I(Y:AC) \leq H(C|AB) \label{eq:mono}
\end{equation}
and permutations of the triples $(X,A,B), (Y,C,A), (Z,B,C)$. To prove this inequality, let's  take the sum of the following {non-negative} entropic quantities:
$I(X:Y|ABCZ)$, $I(Y:Z|ABC)$, $I(X:Z|ABC)$, $I(X:A|BC)$, $I(Y:B|AC)$, $I(Z:C|AB)$, $H(ABCXYZ)- H(X)-H(Y)-H(ABZ)$. Rewriting this sum leads to~\eqref{eq:mono}. Note that {non-negativity} of the last quantity follows from $H(C|ABXYZ)\geq 0$ and $H(ABXYZ)=H(ABZ)+H(X)+H(Y)$, which must hold for any distribution in this configuration where the $X$, $Y$, $Z$ are chosen independently (due to the relativistic causality constraint $P(ab|xyz)=P(ab|z)$ \chanfin{ $\forall \ a,b,x,y,z$}). For a systematic method to derive further such entropy inequalities, we refer to {the Methods}.

Intuitively, \eqref{eq:mono} captures that if $X$ and $Y$ hold a lot of information about correlations of the pairs $B,C$ and $A,C$ respectively, then this restricts the possible dependency of $C$ on $AB$ (which is highest if $H(C|AB)=0$). Intuitively, this has to be the case, since otherwise there would be information about $X$ and $Y$ also contained in $C$, which is not allowed in this setup.

While the triangular arrangement of Figure~\ref{fig:structure} exhibits an appealing symmetry, an immediate question is whether this type of relation also arises in a smaller {setup}. Considering just one of the variables $X$, $Y$, $Z$ cannot lead to similar {relations}, however, dropping only one of them can lead to similar effects. We treat such an example in detail in { Supplementary Note 2},  
where we classify all correlations. There, we furthermore provide examples of correlations, where both $X$ and $Y$ have an effect on the correlations of $A$, $B$, $C$ (but in a weaker form due to the monogamy relations). We further identify cases where $X$ and $Y$ only jointly affect the correlations of $A$, $B$, $C$.

\subsection{Consequences of jamming} \label{sec:signalling}

Our monogamy relations show that relativistically causal correlations (and thus also potential mechanisms leading to them) can be highly non-local. Here, we illustrate this with an example. We show that while the correlations themselves do not (by definition) feature a superluminal transfer of information, { inconsistencies that could be exploited to send superluminal signals} can occur if we assume that there is a mechanism that an agent can { voluntarily} turn on { and off} that { affects} these relativistically causal effects. We call this a \emph{jamming mechanism}~\cite{GPR_jam}. { Specifically, in the following we show that we cannot have the two following features at the same time: (i) a general jamming mechanism that parties can turn on and off (ii) relativistically causal correlations \chanfin{P of $A,B,C,X,Y$}.} 

\begin{figure}
   \centering
   \includegraphics[trim={2.6cm 8.2cm 1.8cm 3.0cm},clip,width=\columnwidth]{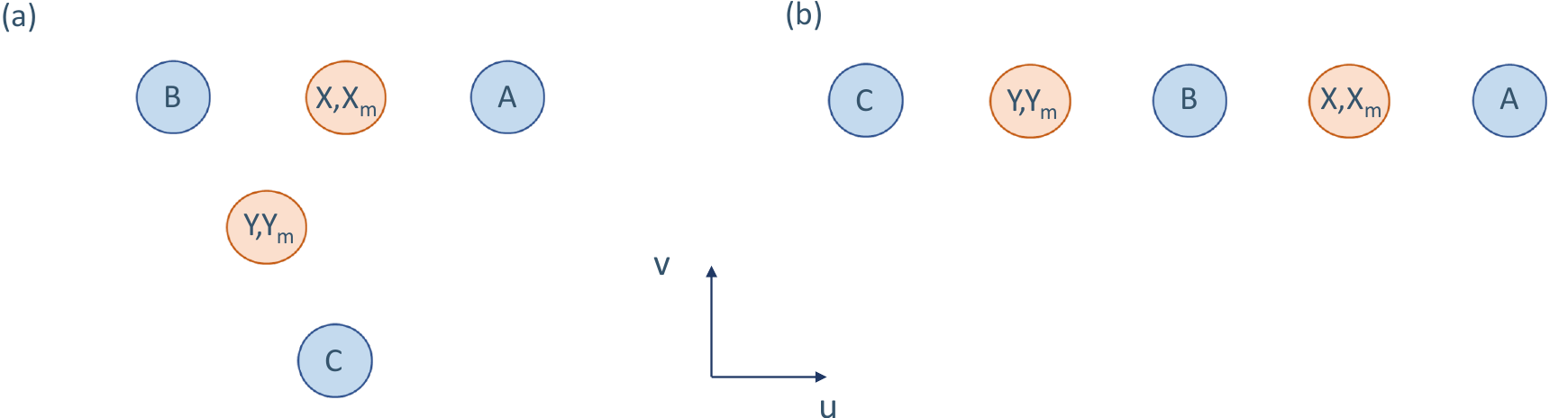}
   \caption{{\bf (a) Compass setup.} The two parties Xavier and { Yanina} (orange) choose inputs { $(X, X_m)$, $(Y, Y_m)$ }while Alice, Bob and Charlie (blue) generate outputs $A$, $B$ and $C$ respectively. The five events are all space-like separated and are displayed here in a frame where they are simultaneous.  
   {\bf (b) Line setup.} This scenario, compared to the compass, does not feature the constraint $P(ac|xy)=P(ac)$ \chanfin{ $\forall \ a,c,x,y$}, while the others remain the same.}  
   \label{fig:compass}
\end{figure}

To see this, let us consider a sub-scenario of the triangular setup above, see Figure~\ref{fig:compass}(a). { We consider three parties,} Alice, Bob and Charlie, who always input 0 into their devices and get outcomes $a,b,c$ respectively and Xavier and Yanina, who choose $(x,x_m)$ and $(y,y_m)$ respectively and obtain an output 0 (where the deterministic inputs and outcomes 0 are not shown in the figure).
Now assume { that the jamming mechanism} that Xavier and Yanina can trigger causes $X$ to influence $A,B$ and $Y$ to influence $B,C$. This is analogous to the scheme in~\cite{GPR_jam}, where a party (e.g.\ Xavier here) can turn on a jamming device that jams the correlations of two others.

Now first of all, { assume that Xavier's jamming mechanism} causes the outputs of Alice and Bob to satisfy $A \oplus B= X$, while without it $A$ and $B$ are independent. Such a mechanism leads to { valid} relativistically causal correlations in { the} sub-scenario that includes $A,B,X$ (as e.g.\ considered in Figure~\ref{fig:bob_jams}).\footnote{In particular, it is straightforward to check that these correlations are allowed \chanfin{in the sub-scenario}.} { If Xavier can turn this mechanism on and off (see (i)), he can  encode the decision to turn his jamming mechanism on as $X_m=1$ or off as $X_m=0$.}  The same strategy can be followed by Yanina for $B,C,Y$ and $Y_m$. 

According to the monogamy relations in this scenario\footnote{This is a sub-scenario of the triangle considered in Section~\ref{sec:monogamy}, where we have dropped one of the sources. In this case, for binary variables $A,B,C$ we are left with the monogamy relation $I(X:AB)+I(Y:BC) \leq 1$. 
}, we know that { for relativistically causal correlations in this setup} whenever $A\oplus B=X$, $B$ and $C$ are independent (and vice versa). { Thus the independent control of the jamming mechanisms (which allows for  $X_m=Y_m=1$) is inconsistent with relativistically causal \chanfin{distribution $P$} (see (ii)).}

This means that { for the distribution \chanfin{$P$} to be consistent with relativistic causality} whenever { Yanina} chooses $Y_m = 1$, Xavier cannot choose $X_m=1$. Thus by choosing $Y_m$ Yanina can { restrict Xavier's choice}: If $Y_m=1$, only $X_m=0$ is possible, if $Y_m=0$ then $X_m \in \{0,1\}$. { The other way around, if we assume that Xavier and Yanina have { control over} a general jamming mechanism { each,} then the overall distribution \chanfin{$P$} will be outside of the relativistically causal set and thus  signalling. Thus such a controlled jamming mechanism could be used to signal.\footnote{Note that there are signalling distributions that are consistent with the two jamming mechanisms:  for each pair $X_m=x_m, Y_m=y_m$ we can construct a (potentially signalling) distribution \chanfin{$P$}. For example, for $X_m=1, Y_m =1$ we can \chanfin{consider} the distribution that has $P(000|00)=P(111|00)=P(010|11)=P(101|11)=P(001|01)=P(110|01)=P(011|10)=P(100|10)=1 \slash 2$ and zero otherwise.}}
{
The simplest way to see this is to consider the case where $X_m=Y_m=1$, which implies that $X= Y\oplus A \oplus C$. Thus Xavier can signal to Alice, Charlie and Yanina, who can use} \chanfin{$A,C,Y$ to retrieve Xavier's} { bit $X$ outside of his future light cone.} 

Now let us assume that {  the jamming mechanism} is only partially controlled, namely, with probability $p>0$ the mechanism of { Yanina} fails, i.e., if $Y_m=1$, $P(B \oplus C = Y)=1-\frac{p}{2}$ and $P(B \oplus C \neq Y)= \frac{p}{2}$, otherwise $B,C$ are independent (and similarly for Xavier). This corresponds to a situation where the parties can only \emph{influence} such a mechanism. Then if $Y_m=1$ this implies that $P(A\oplus B=X) \leq \frac{1}{2}(1-p) + p$. (If $Y_m=0$, then $P(A\oplus B=X) \leq 1$.) Now if Xavier always chooses $X_m=1$, then $P(A\oplus B=X)=1-p+ \frac{1}{2} p$. This implies $p \geq \frac{1}{2}$.

Thus, turning on (or not) a jamming device (e.g.\ $Y_m$) can in this scenario \emph{not} strongly affect whether jamming occurs (e.g.\ $Y=B\oplus C$). Instead, this \chanfin{could potentially be achieved by some} non-local mechanism that takes all potential parties that may attempt to jam into account. We discuss this further in the following.

\subsubsection{Further challenges for establishing { an explanation of} relativistically causal correlations}

\chanfin{We first address the question}, { from which point on information about \chanfin{a jamming  effect could even exist. We illustrate this with the example of the influence of $X$ and/or $Y$ on $B$.} }

\chanfin{Firstly,} 
relativistically causal influences \chanfin{might} take shape where $B$ is generated (or at one specific point in the intersection of the future light cones of $X$ and $Y$ before that).  
However, if a \chanfin{superluminal} influence from $X$ and $Y$ \chanfin{exists only from $B$ on}, (and similarly for $A$, $C$) there needs to be a feature in that { explanation} that causes the scenarios in Figure~\ref{fig:compass} to allow different correlations.
{Thus (non-local) information related to the positions of the other parties would need to be available at $B$ as well.} 

\chanfin{Secondly,}  
 whether $B$ is affected by the superluminal influences of $X$ and/or $Y$ might be \chanfin{determined} where the future light cones of $X$ and $Y$ intersect. However, since for more than one spatial dimension the intersection of the future light cones of  $X$ and $Y$ is not a light cone and does thus not have a single point where the information can start to exist, this  would be a delocalised region.  
\chanfin{This means that along this whole region} there would be a (conspiratorial) decision whether the effect of $X$ or $Y$ should prevail: \chanfin{also a rather non-local mechanism would be needed to ensure this.}

\chanfin{Thirdly,} the jamming variables $X,Y$ might be determined in a conspiratorial way, so that it is already decided which one is to affect $A$, $B$, $C$ when they are generated, i.e., with a non-local mechanism. However, this questions the whole approach of considering light cones, as it seems to make more sense then to consider a single hidden source affecting all variables { including the inputs that should be independent}, as the underlying mechanism that explains the correlations is such. This is in contrast with the relativistic causality framework, which relies on the premise that there are independent variables (here $X,Y$).

{ Any mechanisms leading to such correlations thus has to be highly non-local} \chanfin{in} a stronger \chanfin{sense} than is observed from monogamy relations in quantum theory, e.g.\ for entanglement~\cite{entanglement_monogamy} or non-locality~\cite{nonlocality_monogamy}. In the case of entanglement and non-locality, the monogamy concerns measurement on a system that is shared with several parties that measure it in the future, in the present type of monogamy it concerns  the influences of variables ($X,Y$) that are independent and space-like separated.

\subsubsection{Challenging the possibility of causal inferences} \label{sec:causal}
Causal reasoning in the Bayesian sense~\cite{Pearl} has been successfully adapted from its classical formulation for random variables to the quantum realm~\cite{barrett_framework_1, barrett_framework_2}, where notions such as d-separation and the notion of interventions were formulated for quantum processes. 
Since quantum theory may one-day be replaced by a more general theory (hopefully including gravity), a more general, { theory-independent} framework to study causal relations has been developed for generalised probabilistic theories~\cite{Henson} -- which (as mentioned above) satisfy all non-signalling constraints{~\cite{Masanes}} -- as well as more generally~\cite{VilasiniRogerPRA}. This latter framework is in particular able to incorporate causal influences as considered in this work, allowing for \emph{fine-tuned causal relations}~\cite{VilasiniRogerPRA}.  
 \chanfin{The} monogamy relations considered here, however, provide a conceptual problem for causal reasoning. Causal modelling presupposes that there is an underlying mechanism for correlations. 
Our work suggests that such mechanisms have likely a global nature and can not be (partially) controlled by agents. This questions whether causal modelling makes sense for such correlations in {setups in} more than one spatial dimension, i.e., whether~\cite{VilasiniRogerPRA} can be generalised to this regime.  

Furthermore, by looking at a sub-scenario of the one above, say $A, B, X$, we can potentially not know whether the variables have to follow the non-signalling constraints { (see Supplementary Equation~(S1))} or whether they are only restricted by~\eqref{eq:jam}, which, according to the monogamy relations, depends on { other variables (in this case e.g. $Y$)}. Recall that, since we can { consider} additional inputs to $A,B$ and outcomes to $X$, this { reasoning} also {applies to} the original scenario of Figure~\ref{fig:bob_jams}. This makes causal and also experimental reasoning about setups that allow \chanfin{for general relativistically causal} correlations challenging, in the sense that we have to be able to exclude that there exist any variables (\chanfin{in principle} anywhere in the universe) that may affect correlations \chanfin{e.g.} of $A$ or $B$ with other variables. 
Thus considering such relativistically causal correlations would pose new challenges to the way we currently do research by considering only an experimental setup of interest and some environmental interactions, but where we can assume that no unknown space-like separated variables influence how we can operate our experiment.

\subsection{Further applications of entropy inequalities for relativistic causality}
\label{sec:crypto}
Entropy inequalities further give us a method to distinguish properties of non-signalling from relativistically causal correlations. We show in this section that the latter correlations still satisfy non-trivial constraints that are albeit weaker than those for correlations satisfying the full non-signalling constraints, which were derived in~\cite{BudroniChaves}.

That we can distinguish the correlations of different theories (e.g.\ classical and quantum) in networks with entropy inequalities, has only ever been shown for entropies where we condition on the inputs taking specific values~\cite{BraunsteinCaves, Chaves2012, entropy_review}\footnote{This is called the post-selection technique in~\cite{entropy_review}.}. This motivates us to also pursue such an approach for relativistically causal correlations. 
Our method to obtain constraints for relativistically causal correlations is similar to the method for constraining full non-signalling correlations in~\cite{BudroniChaves}, to which we compare our constraints.  
We describe this in detail in {the Methods.}

To illustrate the technique we show how conditions on the monogamy of non-locality in the scenario of Figure~\ref{fig:bob_jams} have to be relaxed in the relativistically causal case. \chanfin{Let us} stress here that this is a conceptually different type of monogamy relation from those of Section~\ref{sec:monogamy}, the fact that { both} are known as {monogamy} relations is coincidental. Previously, it was shown in~\cite{HR_jam} that the monogamy of the CHSH violation of Alice and Bob and Bob and Charlie is violated by some relativistically causal  correlations. Here we address this { same} problem from the perspective of entropy inequalities, where we show how such inequalities are relaxed as compared to the full non-signalling setting.

For this purpose, let us consider the entropy of the distributions $P(ABC|X=x,Y=y,Z=z)$, where $H(A_xB_yC_z)=H(P(ABC|X=x,Y=y,Z=z))$ is the Shannon entropy of the conditional distribution. Now let us define
\begin{equation}
H^{\rm CHSH}_{AB}:=H(B_0|A_0)+H(A_1|B_0)+H(A_0|B_1)-H(A_1|B_1),   \label{eq:chsh} 
\end{equation} 
where $H(S|T)= H(ST)-H(T)$ is the conditional entropy of the random variable $S$ conditioned on $T$ (according to~\cite{BraunsteinCaves}, $H^{\rm CHSH}_{AB}$ is positive for local models in the Bell scenario). Now, according to~\cite{BudroniChaves}, in the setting of three parties on a line,
\begin{equation}
  H^{\rm CHSH}_{AB}+H^{\rm CHSH}_{BC} \geq 0,  \label{eq:abbc} 
\end{equation} 
which is an entropic version of the monogamy of the non-local correlations Bob holds with Alice and Charlie respectively. In the full non-signalling case this also holds for all permutations of the three parties and relabellings of inputs. 

Following our { technique} detailed in { the Methods},  
we find that when Bob's input can affect the correlations of Alice and Charlie, then the entropic monogamy relation \eqref{eq:abbc} breaks down and we are unable to derive a weaker form of it. 
However, for the correlations Alice shares with Bob and with Charlie there is still a non-trivial, if weaker, monogamy relation:
\begin{align}
H^{\rm CHSH}_{AB}+H(C_0|A_0 Y=1) +H(A_1|C_0 Y=1) \nonumber \\ +H(A_0|C_1 Y=0)- H(A_1|C_1 Y=0) \geq 0, \label{eq:weak_monogamy}
\end{align}
where the second to fifth term reduce to  $H^{\rm CHSH}_{AC}$ when $Y$ does not affect the correlations of $A$ and $C$. For a general technique to derive entropy inequalities for relativistically causal correlations and the derivation of \eqref{eq:weak_monogamy} we refer to { the Methods.}  
Note  also that in this setting the role of the two parties attempting to implement a protocol (here Alice and Bob) is not symmetric anymore.

\section{Conclusions} \label{sec:conclusion}
This work shows that relativistically causal correlations have to satisfy strong monogamy relations, which come about when multi-partite relativistically causal correlations in more than one spatial dimension are to be considered. Such scenarios arguably have to be included in the treatment if  {one} is to take such correlations seriously. 

This puts into question the explanation of these correlations via a jamming mechanism in terms of a device that an agent can voluntarily turn on, as such a device could be used to signal.
This is problematic \emph{in any single reference frame} \chanfin{where} an explanation of these correlations may be attempted, not only when changing reference frames (which is known to lead to difficulties for explaining this type of correlation~\cite{HR_jam}). 
{ These monogamy relations are} also problematic for the way we perform experiments, where we can usually assume that systems at far away space-time locations do not influence the correlations of our outcomes. 

This lack of control about a jamming mechanism also implies that it is not so obvious what the implications of relativistically causal correlations may be for cryptography, since it is not so clear to what extent anyone could actively make use of them. { The monogamy relations further imply} that, if there exists a mechanism that consistently jams the correlations of some parties, this may shield other parties from similar interferences, which may serve as an asset for salvaging some cryptographic protocols (of which several have already been shown to be compromised in this framework in~\cite{HR_jam, HR_jam2}). 

\section{Methods} \label{sec:methods}

\subsection{A method for deriving entropy inequalities for relativistically causal correlations} \label{app:entropy_method}
Entropy allows us to formulate natural measures for quantifying the information some random variables hold about others, in terms of (conditional) mutual information. They furthermore have the convenient property that they linearise (conditional) independence constraints, which has made them one of the main tools to formulate causal compatibility constraints in networks in the past~\cite{Pearl}. 

In a similar vein, any relativistic causality constraint { according to Definition~\ref{def:rel_caus},} leads {also} to an entropic constraint on 
$P(a_1, \ldots, a_n, x_1, \ldots, x_n)$, namely 
\begin{equation}
  H(\tilde{\mathcal{A}}_\mathcal{J} \tilde{\mathcal{X}}_{\mathcal{J}} \tilde{\mathcal{X}}_{\mathcal{J}^c}   )= H(\tilde{\mathcal{A}}_\mathcal{J} \tilde{\mathcal{X}}_{\mathcal{J}}  \tilde{\mathcal{X}}_\mathcal{K})+H(\tilde{\mathcal{X}}_{\mathcal{J}^c} \setminus \tilde{\mathcal{X}}_\mathcal{K}), \label{eq:caus}  
\end{equation}
{where $\tilde{\mathcal{A}}_\mathcal{J}=\{A_i \mid i \in \mathcal{J} \}$ and $\tilde{\mathcal{X}}_\mathcal{J}=\{X_i \mid i \in \mathcal{J} \}$ and analogously for $\tilde{\mathcal{X}}_\mathcal{K},\tilde{\mathcal{X}}_{\mathcal{J}^c}$.}
{ This} can be concisely written as $I(\tilde{\mathcal{A}}_\mathcal{J} \tilde{\mathcal{X}}_{\mathcal{J}}  \tilde{\mathcal{X}}_\mathcal{K}:\tilde{\mathcal{X}}_{\mathcal{J}^c} \setminus \tilde{\mathcal{X}}_\mathcal{K})=0$. It is well known that this holds iff $P(\tilde{\mathcal{A}}_\mathcal{J} \tilde{\mathcal{X}}_{\mathcal{J}} \tilde{\mathcal{X}}_{\mathcal{J}^c} )=P(\tilde{\mathcal{A}}_\mathcal{J} \tilde{\mathcal{X}}_{\mathcal{J}}  \tilde{\mathcal{X}}_\mathcal{K})P(\tilde{\mathcal{X}}_{\mathcal{J}^c} \setminus \tilde{\mathcal{X}}_\mathcal{K})$. Thus, these entropic relations fully capture the relativistic causality constraints.
Recall that as inputs the $X_i$ are {furthermore} all independent, { i.e., $H(X_1, \ldots X_n)=H(X_1)+ \cdots +H(X_n)$}. 

All such constraints, together with the so-called \emph{Shannon inequalities} 
\begin{align}
H(\mathcal{T}) \geq 0 \nonumber \\
H(\mathcal{V} | \mathcal{W}) \geq 0 \label{eq:sh}\\
I(\mathcal{T} : \mathcal{V}|\mathcal{W}) \geq 0 \nonumber
\end{align}
for all non-intersecting sets of variables $\mathcal{T},\mathcal{V},\mathcal{W} \subset \{ A_1, \ldots, A_n, X_1, \ldots, X_n \}$, form the 
so-called \emph{Shannon cone of the scenario} 
\begin{align*}
&\mathcal{S}= \left\{ {\bf{H}} \in \mathbb{R}^{2^{2n}-1} | \eqref{eq:sh} \text{ and } \eqref{eq:caus}  \text{ hold} \right\}, \end{align*}
where $${\bf{H}} \! \!= \! \!(H \!(A_1), \! \ldots \!, \! H \!(X_n), \! H \!(A_1 A_2), \! \ldots \!, \! H \!(A_1 \! \cdots \! A_n X_1 \! \cdots \! X_n) ).$$

Asking what these constraints impose on the relation of specific marginals can then be phrased as a variable elimination problem. Indeed we can identify components of ${\bf{H}}$ that are of interest to us and then derive constraints only involving these variables by means of performing a Fourier-Motzkin elimination of all other variables on $\mathcal{S}$. 
We illustrate this technique with an example.

\begin{example}
Let us consider the triangle setup of Figure~\ref{fig:structure}. We consider the entropy vectors
$${\bf{H}}= (H(A), H(B), \ldots, H(Z), H(AB), \ldots, H(ABCXYZ)),$$
for which we impose all Shannon inequalities as well as the following independence constraints:
\begin{align*}
    &H(ABXYZ)=H(ABZ)+H(X)+H(Y), \\ &H(ACXYZ)=H(ACY)+H(X)+H(Z), \\ &H(BCXYZ)=H(BCX)+H(Y)+H(Z), \\ &H(AXYZ)=H(A)+H(X)+H(Y)+H(Z) \\
    &H(BXYZ)=H(B)+H(X)+H(Y)+H(Z), \\ &H(CXYZ)=H(C)+H(X)+H(Y)+H(Z), \\
    &H(ACYZ)=H(ACY)+H(Z), \\ &H(ACXY)=H(ACY)+H(X), \\
    &H(ACXZ)=H(AC)+H(X)+H(Z), \\ &H(ABYZ)=H(ABZ)+H(Y), \\
    &H(ABXZ)=H(ABZ)+H(X), \\
    &H(ABXY)=H(AB)+H(X)+H(Y), \\
    &H(BCXY)=H(BCX)+H(Y), \\
    &H(BCXZ)=H(BCX)+H(Z), \\
    &H(BCYZ)=H(BC)+H(Y)+H(Z).
\end{align*}
In the context of monogamy relations, we are interested in constraints that only involve 
$H(ABZ)$, $H(ACY)$, $H(BCX)$ \chanfin{and their marginals}. Thus we remove all other variables from the inequalities by means of a Fourier-Motzkin elimination, which we perform using the software PORTA~\cite{PORTA}. From this elimination we find six equality constaints:
\begin{align*}
H(AY)=H(A)+H(Y), \quad H(AZ)=H(A)+H(Z) \\
H(BX)=H(B)+H(X), \quad H(BZ)=H(B)+H(Z) \\
H(CX)=H(C)+H(X), \quad H(CY)=H(C)+H(Y)    
\end{align*}
\chanfin{and} one interesting class of inequality, namely,
$$ I(X:BC)+I(Y:AC) \leq \min \{ H(C|B), H(C|A) \},$$
and permutations of the triples $(X,B,C)$, $(Y,C,A)$, $(Z,A, B)$.  All other inequalities are Shannon inequalities.\footnote{This differs slightly from the inequality in the main text, as in the example presented here for illustration,  we are interested in relaxed constraints that do not involve $H(ABC)$.}
\end{example}

We { remark} that bounds on the cardinality $|S|$ of any involved variable $S$, can in principle be incorporated in this method by imposing additional upper bounds\footnote{This is a relaxation of an actual constraint on the cardinality, as there are variables with higher cardinality that also satisfy the constraint.}
$$H(S) \leq \log |S|.$$
In this scenario this did not lead to any interesting new inequalities.

\bigskip

Entropy inequalities can further be formulated where we condition on some of the variables, usually the inputs, to take specific values. Such methods have been called  post-selection techniques~\cite{entropy_review}. Similar ideas have been applied to analysing networks without the use of entropy, where conditioning on certian variables taking specific values is now known as unpacking~\cite{NavascuesWolfe}. 
The method \chanfin{introduced here} is inspired by~\cite{BudroniChaves}, and allows us to directly compare the inequalities \chanfin{for relativistically causal correlations} to those arising from the full non-signalling correlations from~\cite{BudroniChaves}.

For this, let us consider a \chanfin{collection of pairs of} (input, output) variables  $(X_i,A_i)$ that can take values $x_i \in \mathcal{X}_i$ and $a_i \in \mathcal{A}_i$ respectively.  
\chanfin{Then let us consider} a network involving all of these variables with \chanfin{a directed edge from input to outcome within each pair}. 
Now for each ${X}_i$ in the network we identify all variables $A_j$ of which ${X}_i$ \chanfin{may influence} the correlations with some other variables  and we add \chanfin{directed edges} from $X_i$ to all of these variables. After this, we consider each $A_j$ and we create a copy of this variable for each combination of values the variables influencing it can take. For instance, if $A_j$ is influenced in this network by $X_j, X_k, X_l$ (meaning that there is \chanfin{a directed edge} pointing to it from those variables), then we create $|\mathcal{X}_j| \cdot |\mathcal{X}_k| \cdot |\mathcal{X}_l|$ copies of $A_j$, indexed by the values $x_j,x_k,x_l$, namely $A_j(x_j,x_k,x_l)$. These represent random variables with \chanfin{probabilities} $P(a_j|x_j,x_k,x_l)$.

Now among all the variables $\{A_i(\cdot)\}_i$ we define sets of coexisting variables, which are those that agree in the value that the $X_j$ that they share as an index take and we drop all such sets that are subsets of others, thus leaving us with maximal coexisting sets. These are the {maximal} sets of variables for which we can define joint conditional \chanfin{distributions}.
For each of these we build an entropy vector.  

Now, each of these entropy vectors has to satisfy all Shannon inequalities (see above). In addition, the relativistic causality constraints impose relations among the entropies: If a variable, $X_j$, affects \emph{only} the correlations of another, $A_i$, the entropy of $A_i$ will not depend on $X_j$, i.e., $H(A_i(x))=H(A_i({x'}))$ for two values $x,x'$ of $X_j$.\footnote{\chanfin{Analogous relations may arise for sets of variables $\{ A_i \}_i$.}} 
Note furthermore that the entropy vectors share some of the variables, so that we obtain an overall system of linear entropy inequalities by combining the inequalities from all the maximal coexisting sets.

Now we can use Fourier-Motzkin elimination to obtain constraints that only involve the variables we are interested in. We illustrate this whole procedure with an example.

\begin{example}
    Let us consider the scenario from Figure~\ref{fig:bob_jams}, with binary inputs $X,Y,Z$. This means that we define $10$ variables $A_{00}=A_{x=0,y=0}$, $A_{01}=A_{x=0,y=1}$, $A_{10}=A_{x=1,y=0}$, $A_{11}=A_{x=1,y=1}$, $B_0=B_{y=0}$, $B_1=B_{y=1}$, $C_{00}=C_{y=0,z=0}$, $C_{01}=C_{y=0,z=1}$, $C_{10}=C_{y=1,z=0}$, $C_{11}=C_{y=1,z=1}$. These form six maximal coexisting sets $\{A_{x_0y_0}, B_{y_0}, C_{y_0 z_0} \}$ for $x_0, y_0, z_0 \in \{0,1\}$. For each set we obtain a vector $${\bf H}_{x_0,y_0,z_0}=(H(A_{x_0 y_0}), H(B_{y_0}), \ldots, H(A_{x_0 y_0} B_{y_0} C_{y_0z_0})).$$
    In this case, relativistic causality imposes that 
    \begin{align*}
    H(A_{00})=H(A_{01}), \quad H(A_{10})=H(A_{11}) \\
    H(C_{00})=H(C_{10}), \quad H(C_{01})=H(A_{11}).      
    \end{align*} 
    We further impose all Shannon inequalities.
    
    In our example we are interested in the tradeoff between the CHSH violation of Alice and Bob, Alice and Charlie. We thus aim to derive constraints involving $H^{\rm CHSH}_{AB}$, $H^{\rm CHSH}_{AC}$, (see \eqref{eq:chsh} in the main text). For this we remove entropies involving three of the variables as well as those involving $B$ and $C$ from the scenario. Performing this variable elimination in PORTA~\cite{PORTA}, we obtain Shannon inequalities and $16$ inequalities of the type  \eqref{eq:weak_monogamy}. 
\end{example}

\begin{acknowledgements}
The author would like to thank Paul Skrzypczyk and Giorgos Eftaxias for helpful discussions and Roger Colbeck and Peter Brown for feedback on a draft of this work. This work was funded by the Swiss National Science Foundation (Ambizione PZ00P2\textunderscore208779).   
\end{acknowledgements}

\bibliography{jamming_bib.bib}

\begin{appendix}

\onecolumngrid

\renewcommand*{\thesection}{Supplementary Note \arabic{section}}

\renewcommand{\appendixname}{} 

\renewcommand{\theequation}{S\arabic{equation}}

\section{Relativistic causality and the dependency on the number of spatial dimensions} \label{app:example}

{ This section provides more details on the definition of relativistically causal correlations. In Section~\ref{sec:rel_caus}, we give a formal definition of the multipartite non-signalling constraints, which allows for comparison with relativistic causality. In Section~\ref{sec:dimensionality}, we contrast this with relativistically causal constraints, where we illustrate how the valid constraints in a specific setup may depend on the number of spatial dimensions and the arrangement of the involved parties.}

\subsection{Multipartite non-signalling conditions} \label{sec:rel_caus}

{ In a setting involving $n$ spacelike separated parties $1, \ldots n$, who can choose inputs $X_i$ and obtain outcomes $A_i$, the \emph{n-partite non-signalling constraints}}
can be written as~\cite{Masanes}
\begin{align}
 \sum_{a_k} &P( \{a_1, \ldots,  a_n \}| \{x_1, \ldots, x_n \}) \nonumber \\
 &= P( \{a_1, \ldots, a_n \} \setminus a_k | \{x_1, \ldots, x_n \} \setminus x_{k}) \label{eq:ns}
\end{align}
$\forall \ k, \{a_1, \ldots, a_n \} \setminus a_k, x_1, \ldots, x_n$. { For many arrangements of the parties, these constraints differ from the relativistic causality constraints of Definition~II.1 from the main text.}

{\subsection{Relativistically causal constraints in different arrangements and dimensions} \label{sec:dimensionality}
}

{ To illustrate the role of Definition~II.1 of the main text, we illustrate here} that there are constellations of events, where it leads to constraints that are weaker than the ones that one would obtain from~\eqref{eq:ns}. { To see} this for a specific arrangement, it is important to take \emph{all} spatial dimensions into account when identifying whether a variable can potentially affect the correlations of others. 
We illustrate the dimension dependency by means of the example from {Figure~1 from the main text in the following.}

\begin{example}

{ Let us first remark that if we consider the setup from Figure~1 from the main text in a world with just one spatial dimension, then spacelike separation of the three parties is sufficient for 
 the three parties to display  correlations that differ from~\eqref{eq:ns}. 
 
 In more spatial dimensions this is not the case anymore and in the following we show how
 the positions of $(x,a)$, $(y,b)$ and $(z,c)$ in Figure~1
    affect this in different dimensions.}

Let us consider the example from {Figure~1 of the main text} but now take one additional spatial dimensions into account. For convenience of notation we choose units such that the speed of light is 1. 
Thus (in our reference frame from before) we have three events at coordinates $(t,u,v)=(0,0,0)$ where $a$ was generated from $x$, at $(t,u,v)=(0,\frac{r}{2},0)$ where $b$ was generated from $y$ and at $(t,u,v)= (0,r,0)$, where $c$ was generated from $z$. Note that $r >0$.

Now let us consider the boundary of the intersection of the future light cones of $a$ and $c$. This is the intersection of two cones parametrised by $r_a, s_a$ and $r_c, s_c$ as 
\begin{align*}
    \mathcal{K}_a &=\left\{ p_a \mid p_a= \left(t_a, r_a, s_a\right), \ \sqrt{t_a^2-r_a^2} \geq |s_a|, \  t_a \geq |r_a| \right\}, \\
    \mathcal{K}_c &=\left\{ p_c \mid p_c= \left(0,r,0 \right)+\left(t_c, r_c, s_c\right), \ \sqrt{t_c^2-r_c^2} \geq |s_c|, \  t_c \geq |r_c| \right\},
\end{align*}
namely $\mathcal{S}_{ac} = \mathcal{K}_a \cap \mathcal{K}_c$. 
To characterise $\mathcal{S}_{ac}$, we first find the points that are simultaneously on the boundary of $\mathcal{K}_a$ and $\mathcal{K}_c$, namely
$$\tilde{\mathcal{S}}_{ac}= \left\{ p_{ac} \mid p_{ac} = \left(  t_{ac}, \frac{r}{2}, \pm \sqrt{t_{ac}^2-(\frac{r}{2})^2}\right), \  t_{ac} \geq \frac{r}{2} \right\}. $$ Now $\mathcal{S}_{ac}$ is the union of all light cones starting at points in $\tilde{\mathcal{S}}_{ac}$. 
Notice in particular that this set itself is not a light cone, i.e., its points are not within the future of a single point (as would be the case in $1+1$-dimensions). Namely, take $\delta >0$ and compare the two points $(\frac{r}{2}, \frac{r}{2}, 0)$, $(\frac{r}{2}+\delta, \frac{r}{2},  \sqrt{(\frac{r}{2}+\delta)^2-(\frac{r}{2})^2}) \in \tilde{\mathcal{S}}_{ac}$. Now the second point cannot be reached by  
$(\frac{r}{2}, \frac{r}{2}, 0)+ (\delta, 0, s)$, since in a future light cone $s \leq \delta$ but  $\delta < \sqrt{\delta^2+r \delta}$ for $\delta >0$. 

Now let us consider the light cone 
$$\mathcal{K}_b = \left\{ p_b \mid p_b= \left(0,\frac{r}{2}, 0 \right)+\left(t_b, {r_b}, s_b  \right), \ \sqrt{t_b^2-r_b^2} \geq |s_b|, \ t_b \geq |r_b| \right\}. $$
We observe that $\tilde{\mathcal{S}}_{ac} \subset \mathcal{K}_b$, which is directly seen as $r_b=0$ is allowed in $\mathcal{K}_b$ and thus also ${\mathcal{S}}_{ac} \subset \mathcal{K}_b$. 
Notice that since $\mathcal{S}_{ac}$ is the union of light cones that are all within $\mathcal{K}_b$, this will hold in all inertial reference frames. 
However, in contrast to the one dimensional case above, where $b$ could be generated at a point $(\frac{r}{2}, \frac{r}{2}, 0)$, we cannot allow $b$ to be generated at a point $(r_0, \frac{r}{2}, 0)$ with $r_0 > 0$ in the case of two spatial dimensions, since  for ${\mathcal{S}}_{ac} \subset \mathcal{K}_b$ we need that $r_0 \leq  t_{ac} - \sqrt{t_{ac}^2-\left(\frac{r}{2}\right)^2}$ for all $t_{ac} \geq \frac{r}{2}$. A similar argument holds for $r_0 < 0$.

The requirements on the setup in more than one spatial dimension are not only more rigid when considering the time dimension but also when trying to vary the spatial dimensions. Specifically, consider instead a starting position for $b$ that is slightly out of plane, i.e., for some $\epsilon > 0$,
$$\mathcal{K}_b' = \left\{ p_b \mid p_b= \left(0,\frac{r}{2}, -\epsilon \right)+\left(t_b, {r_b}, s_b  \right), \ \sqrt{t_b^2-r_b^2} \geq |s_b|, \  t_b \geq |r_b| \right\}. $$
This implies that 
$\epsilon \leq t_{ac}-\sqrt{t_{ac}^2-(\frac{r}{2})^2} \ \forall \ t_{ac} \geq \frac{r}{2}$. Thus as $t_{ac} \rightarrow \infty$, we observe that $\epsilon \rightarrow 0$.
In the line connecting $a$ and $c$, the point $b$ may however move, i.e., we can have 
$$\mathcal{K}_b'' = \left\{ p_b \mid p_b= \left(0,\frac{r}{2}-\epsilon, \chanfin{0}  \right)+\left(t_b, {r_b}, s_b  \right), \ \sqrt{t_b^2-r_b^2} \geq |s_b|, \ t_b \geq |r_b| \right\}, $$
as long as $-\frac{r}{2} \leq \epsilon \leq \frac{r}{2}$.
\end{example}

A further generalisation to three spatial dimensions proceeds along the same lines. \chanfin{The intersection of two future light cones in three spatial dimensions is a collection of future light cones:} The region where two light cones, 
\begin{align*}
    \mathcal{K}_a^3 &=\left\{ p_a \mid p_a= \left(t_a, r_a, s_a, q_a\right), \ \sqrt{t_a^2-r_a^2-s_a^2} \geq |q_a|, \ \sqrt{t_a^2-r_a^2} \geq |s_a|, \  t_a \geq |r_a| \right\}, \\
    \mathcal{K}_c^3 &=\left\{ p_c \mid p_c= \left(0,r,0,0 \right)+\left(t_c, r_c, s_c, q_c \right), \ \sqrt{t_c^2-r_c^2-s_c^2} \geq |q_c|, \ \sqrt{t_c^2-r_c^2} \geq |s_c|, \ t_c \geq |r_c| \right\},
\end{align*}
intersect is in this case given by the future of any points within
$$\tilde{\mathcal{S}}_{ac}^3= \left\{p_{ac} | p_{ac}= \left(t_{ac}, \frac{r}{2}, s_{ac}, q_{ac} \right), \ t_{ac}^2-(\frac{r}{2})^2-s_{ac}^2-q_{ac}^2 \geq 0 \right\}.$$

In general, the future of a set of variables $\mathcal{A}_\mathcal{J}$ is always a union of future light cones of some set of points $\tilde{\mathcal{S}}_\mathcal{J}$, that is a subset of a light cone $\mathcal{K}_\mathcal{J}$ that in some situations does not contain the full future light cones of all variables in $\mathcal{A}_\mathcal{J}$.
\chanfin{This generalisation proceeds} inductively in the sense that the intersection of a collection of light cones $\cap_i A_i$ is still a collection of light cones. Namely, $A_1 \cap A_2$ is a collection of light cones by the above, furthermore $A_1 \cap A_2 \cap A_3$ is obtained by intersecting each of the light cones \chanfin{in the collection} with that of $A_3$ and then taking the union of the resulting light cones and so forth. 

{
\section{Extremal correlations and monogamy relations} 
}

{ In this section we provide more details on the extremal correlations for the scenarios considered in Figures 2 and 3 of the main text and provide further examples exhibiting monogamy relations.}

\bigskip

\subsection{Correlations in triangular setup with $Z=A \oplus B$}
\label{app:vertex_enumeration}

Notice that all relations imposed by relativistic causality are linear on the conditional distribution (no matter which light cone arrangement we consider). Additional conditions on the correlations like $P(ab|z)=0$ if $Z \neq A \oplus B$, which can also be imposed as linear equalities for \chanfin{$P$}, can thus be added and the respective polytope can be described in terms of its extremal vertices. In this case, this is made up from convex combinations of the two extremal points

$$P_1(abc|xyz)=\begin{cases} \frac{1}{2} & \text{if } z=a \oplus b, \ c=0 \\
    0 & \text{otherwise} .
    \end{cases}$$
$$P_2(abc|xyz)=\begin{cases} \frac{1}{2} & \text{if } z=a \oplus b, \ c=1 \\
    0 & \text{otherwise} .
    \end{cases}$$

Notice that due to the fine-tuned nature needed to enable $Z=A \oplus B$, imposing $P(ab|z)=0$ if $Z \neq A \oplus B$ leads to the same polytope as imposing $P(ab|z)=\frac{1}{2}$ if $Z = A \oplus B$ in combination with the relativistic causality constraints.

\bigskip

Similar considerations can be made for other distributions and in particular when each of the six parties in the setup choose inputs and obtain outcomes. Let us consider, for instance, the example from~\cite{HR_jam} (Supplementary Note 5). There, a distribution $P_{\rm HR}(a_oy_oc_o|a_iy_iz_i)$ with binary inputs (outcomes) $A_i,Y_i,C_i (A_o,Y_o,C_o)$ is defined that is valid in the setting of { Figure~1 of the main text}. The distribution has marginals 
$$P_{\rm HR}(a_oc_o|a_iy_ic_i)=\begin{cases} P_{ \rm perf}(a_oc_o|a_ic_i) & y_i=0 \\
P_{\rm PR}(a_oc_o|a_ic_i) & y_i=1 \end{cases}$$
and $y_o=0$ independently of all other variables, where $P_{ \rm perf}(a_oc_o|a_ic_i)$ is $\frac{1}{2}$ iff $a_o \oplus c_o = 0$ and $P_{\rm PR}(a_oc_o|a_ic_i)$ is a PR-box, i.e., it is $\frac{1}{2}$ iff $a_o \oplus c_o = a_i \cdot c_i$. For $A_o,C_o,Y_o$ and assuming uniformly random $A_i,Y_i,C_i$, 
these correlations are $P(a_oc_o|y_i)=\frac{1}{4} P_1(a_oc_o|y_i)+ \frac{3}{4} P_2(a_oc_o|y_i)$ where $P_1(a_oc_o|y_i)=\frac{1}{2}$ iff $a_o\oplus c_o = y_i $ and $P_2(a_oc_o|y_i)=\frac{1}{2}$ iff $a_o\oplus c_o = 0 $. 
Now we can ask whether such a distribution could also be implemented in the triangular setup of { Figure~2 of the main text}, with additional inputs for $A_i,B_i,C_i$ and outcomes $X_o,Y_o,Z_o$. 
Thus, now interpreting $A_o,C_o,Y_i$ as the respective variables in this setup, we can see that this type of influence from $Y_i$ onto $A_o,C_o$ is possible. 
A similar treatment as in the previous example,\footnote{Notice that $P(a_oc_o|x_iy_iz_i)=P(a_oc_o|y_i)$, where $X_i$, $Z_i$ are the variables in the triangular setup.}
namely, where we here require that for each $x_i,z_i$ if $y_i=0$, $P(01|y_i)=P(01|y_i)=0$ and if $y_i=1$, $3P(10|y_i)=P(00|y_i)$, $3P(01|y_i)=P(11|y_i)$ and $P(00|y_i)=P(11|y_i)$. Together with the relativistic causality constraints this forms a  polytope of 65 extremal vertices that make up all distributions compatible with the triangle setup and with this marginal $P(a_oc_o|y_i)$. 
We can now check that there is no compatible distribution that has the same marginal for the variables $P(a_ob_o|z_i)$ or $P(b_oc_o|x_i)$, thus the same correlations cannot be implemented in the triangle (with additional inputs to $A,B,C$) along several sides at the same time. We do this by imposing these additional constraints and showing that there is no feasible solution to the resulting linear program. In order not to rely on linear programming, we further compute all possible marginals that we can observe when fixing $P(a_0 c_0|y_i)$ accordingly and find indeed that the same marginals are not possible for $P(a_ob_o|z_i)$ or $P(b_oc_o|x_i)$ under these constraints.

\bigskip

Another relevant distribution in the setting of { Figure~1 of the main text} is the one that was shown in~\cite{HR_jam} to violate  monogamy of CHSH violation (see also {main text} for more on this). This distribution is defined as~\cite{HR_jam} 
\begin{align*}
&P(001|011)=P(110|011)=\frac{1}{2}, \quad P(011|110)=P(100|110)=\frac{1}{2}, \quad P(010|111)=P(101|111)=\frac{1}{2}, \\ &P(000|xyz)=P(111|xyz)=\frac{1}{2} \quad \forall (x_i,y_i,z_i) \notin \{(0,1,1), (1,1,0),(1,1,1) \},
\end{align*} 
all other probabilities are zero.
For uniform $X_i,Y_i,Z_i$ we can show that for this distribution $P(a_oc_o|y_i)=\frac{1}{2} P_1(a_oc_o|y_i)+\frac{1}{2} P_2(a_oc_o|y_i)$. Thus we impose that for each $x_i,z_i$ if $y_i=0$, $P(01|y_i)=P(01|y_i)=0$ and if $y_i=1$, $P(10|y_i)=P(00|y_i)=P(01|y_i)=P(11|y_i)$. In this case we obtain a polytope with $23$ extremal vertices compatible with relativistic causality and these constraints. We further find again that the same marginal is not possible for $P(a_ob_o|z_i)$ or $P(b_oc_o|x_i)$ under these constraints. We check this again with a linear feasibility problem as well as by computing all possible marginals we can have for $P(a_ob_o|z_i)$ or $P(b_oc_o|x_i)$ in this case.

This means that these correlations can only be realised among three parties in the triangle setup of { Figure~2 of the main text}. At the same time having such a relativistically causal influence prevents the other parties from sharing such a distribution and in fact any distribution with such relativistically causal influence.

 \subsection{Full analysis of the compass setup} \label{app:vertex_enumeration_compass}

In this section we give a full description of the scenario displayed in {Figure~3(a) of the main text}. 
The relativistic causality constraints for this scenario are 
\begin{align}
	P(ab|xy) &= P(ab|x) \nonumber\\
	P(ac|xy) &= P(ac) \nonumber\\
	P(bc|xy) &= P(bc|y) \nonumber\\
	P(a|xy) &= 	P(a) \nonumber\\
	P(b|xy) &= 	P(b) \nonumber\\
	P(c|xy) &= 	P(c)  \nonumber
\end{align}
\chanfin{$ \forall \  a,b,c,x,y $}. In this case we obtain only $82$ extremal correlations (again obtained using PORTA~\cite{PORTA} for the vertex enumeration). These come from 6 classes, the elements of which are equivalent up to exchange of $A,X$ with $C,Y$ and up to relabelling of inputs and outcomes. In the following we give an example from each class in functional form:
\begin{itemize}
    \item Deterministic distributions (8 instances): $$P(abc|xy)=\begin{cases} 1 & \text{if } a=0, b=0, c=0 \\
    0 & \text{otherwise} .
    \end{cases}$$
    \item  3-party PR-box-like correlations (10 instances):
    $$P(abc|xy)=\begin{cases} \frac{1}{4} & \text{if } a \oplus b \oplus c = x \cdot y \\
    0 & \text{otherwise} .
    \end{cases}$$
    This example is somewhat curious, since, while $X$ doesn't affect the correlations between $A,B$ and $Y$ doesn't affect the correlations between $B,C$, $X$ and $Y$ \emph{jointly} affect the correlations of $A,B,C$. This is indeed compatible with the space time setup in that the joint future of $A,B,C$ is in the future of the joint future of $X,Y$.

    \item Monogamous correlations (8 instances):
     $$P(abc|xy)=\begin{cases} \frac{1}{2} & \text{if } a \oplus b  = x , \ c=0 \\
    0 & \text{otherwise} .
    \end{cases}$$

     \item  Type IV correlations (16 instances):
     $$P(abc|xy)=\begin{cases} \frac{1}{4} & \text{if } a \oplus b  = (y \oplus c) \cdot x , \ c=0 \\
    0 & \text{otherwise} .
    \end{cases}$$

     \item  Type V correlations (8 instances): 
     $$P(abc|xy)=\begin{cases} \frac{1}{4} & \text{if } x= a \oplus b , y= c \oplus b \ \text{ or } \ x= a \oplus b, y= c \ \text{ or } \ x= a , y= c \oplus b \\
    0 & \text{otherwise} .
    \end{cases}$$
    \item  Type VI correlations (32 instances): 
     \begin{align*}
     P(001|00)=P(010|00)=P(111|00)=0, \ 
     P(000|00)=P(100|00)=P(110|00)=\frac{1}{7}, \ 
     P(011|00)=P(101|00)=\frac{2}{7}, \  \\
     P(000|01)=P(111|01)=0, \ 
     P(001|01)=P(010|01)=P(011|01)=P(100|01)=P(110|01)=\frac{1}{7}, \ P(101|01)=\frac{2}{7}, \  \\
     P(000|10)=P(011|10)=P(101|10)=P(110|10)=0, \ 
     P(010|10)=\frac{1}{7}, \ 
     P(001|10)=P(100|10)=P(111|10)=\frac{2}{7}, \  \\
     P(010|11)=P(100|11)=P(111|11)=0, \ 
     P(000|11)=P(001|11)=P(011|11)=\frac{1}{7}, \ 
     P(101|11)=P(110|11)=\frac{2}{7}. \
     \end{align*}
\end{itemize}

\end{appendix}

\end{document}